\documentclass[epj,twocolumn]{webofc}
\usepackage[varg]{txfonts}   % Web of Conferences font
\woctitle{NSRT15}
\begin{document}
\title{Resonances and Reactions from Mean-Field Dynamics}
%
% subtitle is optionnal
%
%%%\subtitle{Do you have a subtitle?\\ If so, write it here}

\author{P. D. Stevenson\inst{1}\fnsep\thanks{\email{p.stevenson@surrey.ac.uk}}}

\institute{Department of Physics, University of Surrey, Guildford, GU2 7XH, U.K.}

\abstract{%
The time-dependent version of nuclear density functional theory, using functionals derived from Skyrme interactions, is able to approximately describe nuclear dynamics.  We present time-dependent results of calculations of dipole resonances, concentrating on excitations of valence neutrons against a proton plus neutron core in the neutron-rich doubly-magic $^{132}$Sn nucleus, and results of collision dynamics, highlighting potential routes to ternary fusion, with the example of a collision of $^{48}$Ca+$^{48}$Ca+$^{208}$Pb resulting in a compound nucleus of element 120 stable against immiedate fission
}
\maketitle
\section{Introduction}
\label{intro}
The time-dependent version of nuclear mean-field dynamics, usually called time-dependent Hartree-Fock (TDHF) \cite{Sim12}, is able to give a description of low-energy nuclear dynamic processes, such as fusion and quasi-fission \cite{Toh02,Sim04,Uma06,Sim07,Sim08,Sim08b,Uma08,Uma10,Uma12,Ste15,Uma15}, fission \cite{Neg78,Sim14,God15} and collective excitations \cite{Str79,Uma86,Cho87,Sim03,Ste04,Nak05,Uma05,Alm05,Ste07,Sim09,Fra12,Kla13,Ste14}.

The TDHF equations were first written down by Dirac in 1930 \cite{Dir30} with the first quasi-realistic calculations performed in the 70s and 80s \cite{Neg82,Dav85}, with more realistic calculations coming as symmetry assumptions were lifted and more sophisticated effective interactions were used.  Now, codes with no symmetry restrictions and featuring modern forms of the Skyrme interaction \cite{Ben03} are available \cite{Mar14}

In this proceeding we give a brief outline of the method and the physics input, and show applications of TDHF to low-lying modes and ternary fusion reactions.

\section{Theory}
The time-dependent Hartree-Fock equations are a set of coupled non-linear Schr\"odinger equations
\begin{equation}
  i\hbar\frac{d\psi_\alpha(x,t)}{dt} = \hat{h}\psi_\alpha(x,t) \label{eq:tdhf}
\end{equation}
Here. the one-body Hamiltonian $\hat{h}$(t) is shown to depend explicitly on time.  This comes about through its dependence on the densities, which are built up from the time-dependent wave functions.

Assuming that one has a suitable initial set of wave functions, obtained from a static Hartree-Fock calculation, augmented by a suitable boost,
one solves the TDHF equations (\ref{eq:tdhf}) by iterating forward in time with small ($\Delta t\simeq 0.2$ fm/c) time steps through \cite{Mar14}

\begin{equation}
\psi_\alpha(t+\Delta t) = \exp(i\hat{h}_\alpha((t+\Delta t)/2)\Delta t/\hbar)\psi_\alpha(t),
\end{equation}
where the Hamiltonian at the half time step is approximated by using a similar evolution equation with $\Delta t\rightarrow \Delta t/2$ in the exponential and the Hamiltonian at the time $t$.

The boost, typically given as an instantaneous action upon the initial static wave functions, is of a form such as to superimpose a velocity field on the nuclear system, consisting of one or more nuclei.  Collective excitations are initiated by having e.g. a spatially-dependent velocity field acting on a single nucleus while collisions are initialised by given fixed velocity boosts to spacially-separated nuclei.  Details of specific boosts used in this proceeding are given in the sections discussing the specific calculations.

The Skyrme interaction can most conveniently be written as an energy density functional.  This, for quite a general class of forces found in the literature (which can differ in which terms are included), can be written as \cite{Les07}

\begin{eqnarray}
  &&\mathcal{E}_\mathrm{Skyrme} = \nonumber \\
  &&\int d^3r\sum_{t=0,1}\Bigg\{C_t^\rho[\rho_0]\rho_t^2+C_t^s[\rho_0]\boldsymbol{s}_t^2+C_t^{\Delta\rho}\rho_t\nabla^2\rho_t\nonumber \\
&+&C_t^{\nabla s}(\nabla\cdot\boldsymbol{s})^2+C_t^{\Delta s}\boldsymbol{s}_t\cdot\nabla^2\boldsymbol{s}_t+C_t^{\tau}(\rho_t\tau_t-\boldsymbol{j}_t^2) \nonumber \\
&+&C_t^T\left(\boldsymbol{s}_t\cdot\boldsymbol{T}_t-\sum_{\mu,\nu=x}^zJ_{t,\mu\nu}J_{t,\mu\nu}\right) \nonumber \\
&+&C_t^F\Bigg[\boldsymbol{s}_t\cdot\boldsymbol{F}_t-\frac{1}{2}\left(\sum_{\mu=x}^zJ_{t,\mu\mu}\right)^2-\frac{1}{2}\sum_{\mu,\nu=x}^zJ_{t,\mu\nu}J_{t,\nu\mu}\Bigg]\nonumber\\
&+&C_t^{\nabla\cdot J}\left(\rho_t\nabla\cdot\boldsymbol{J}_t+\boldsymbol{s}_t\cdot\nabla\times\boldsymbol{j}_t\right)\Bigg\}.
\label{eq:edens}
\end{eqnarray}

As well as these terms arising from the Skyrme interaction itself, the kinetic energy and Coulomb interaction is included.  The usual Slater exchange approximation is made for the Coulomb force.

In (\ref{eq:edens}), the various $C$ coefficients are, in our case, related to parameters of the original Skyrme interaction, though one may take the alternative approach that the energy density functional (\ref{eq:edens}) is the fundamental expression of the nuclear interaction and directly fit the $C$ coefficients to data.

The various densities and currents are given as

\begin{eqnarray}
\rho_q(\boldsymbol{r})&=&\left.\rho_q(\boldsymbol{r},\boldsymbol{r}')\right|_{\boldsymbol{r}=\boldsymbol{r}'}\nonumber \\
\boldsymbol{s}_q(\boldsymbol{r}) &=& \left.\boldsymbol{s}_q(\boldsymbol{r},\boldsymbol{r}')\right|_{\boldsymbol{r}=\boldsymbol{r}'}\nonumber \\
\tau_q(\boldsymbol{r}) &=& \left.\nabla\cdot\nabla'\rho_q(\boldsymbol{r},\boldsymbol{r}')\right|_{\boldsymbol{r}=\boldsymbol{r}'} \nonumber \\
T_{q,\mu}(\boldsymbol{r}) &=& \left.\nabla\cdot\nabla's_{q,\mu}(\boldsymbol{r},\boldsymbol{r}')\right|_{\boldsymbol{r}=\boldsymbol{r}'}\\
\boldsymbol{j}_q(\boldsymbol{r}) &=& \left.-\frac{i}{2}(\nabla-\nabla')\rho_q(\boldsymbol{r},\boldsymbol{r}')\right|_{\boldsymbol{r}=\boldsymbol{r}'}\nonumber \\
J_{q,\mu\nu}(\boldsymbol{r})&=&\left.-\frac{i}{2}(\nabla_\mu-\nabla_\mu')s_{q,\nu}(\boldsymbol{r},\boldsymbol{r}')\right|_{\boldsymbol{r}=\boldsymbol{r}'}\nonumber\\
F_{q,\mu}(\boldsymbol{r})&=&\left.\frac{1}{2}\sum_{\nu=x}^z(\nabla_\mu\nabla_\nu'+\nabla_\mu'\nabla_\nu)s_{q,\nu}(\boldsymbol{r},\boldsymbol{r}')\right|_{\boldsymbol{r}=\boldsymbol{r}'}\nonumber.
\end{eqnarray}
Here, the subscript $q$ gives the particle type (proton or neutron).  Greek letter subscripts indicate Cartesian coordinates for vector or tensor quantities.  From these densities one then defines the isoscalar
($t=0$) and isovector ($t=1$) densities and currents found in (\ref{eq:edens}) as 
\begin{eqnarray}
\rho_0(\boldsymbol{r}) &=& \rho_n(\boldsymbol{r})+\rho_p(\boldsymbol{r}) \nonumber \\
\rho_1(\boldsymbol{r}) &=& \rho_n(\boldsymbol{r})-\rho_p(\boldsymbol{r}),
\end{eqnarray}
and similarly for the other densities and currents.

\section{Response in $^{132}$Sn}

We performed dipole response calculations in $^{132}$Sn using the KDE0v1 functional \cite{Agr05}, chosen since it has been fitted to some giant resonance (breathing mode) data, and for its good general nuclear matter properties \cite{Dut12}.  The neutron-rich nucleus $^{132}$Sn has had its E1 strength measured, with evidence of a low-lying peak found at 9.8 MeV, while the main GDR peak is found around 17 MeV \cite{Adr05}.

Starting from the ground state, we apply a dipole boost of the following form \cite{God13}:
\begin{equation}
  {D} = \frac{A-\sum_{i=1}^\eta v_i^2}{A}\sum_{i=1}^\eta{x}_i - \frac{\sum_{i=1}^\eta v_i^2}{A}\sum_{j=\eta+1}^{A}{x}_j.\label{eq:boost}
  \end{equation}

\begin{figure}[!tbhp]
  \centering
  \includegraphics[width=8cm]{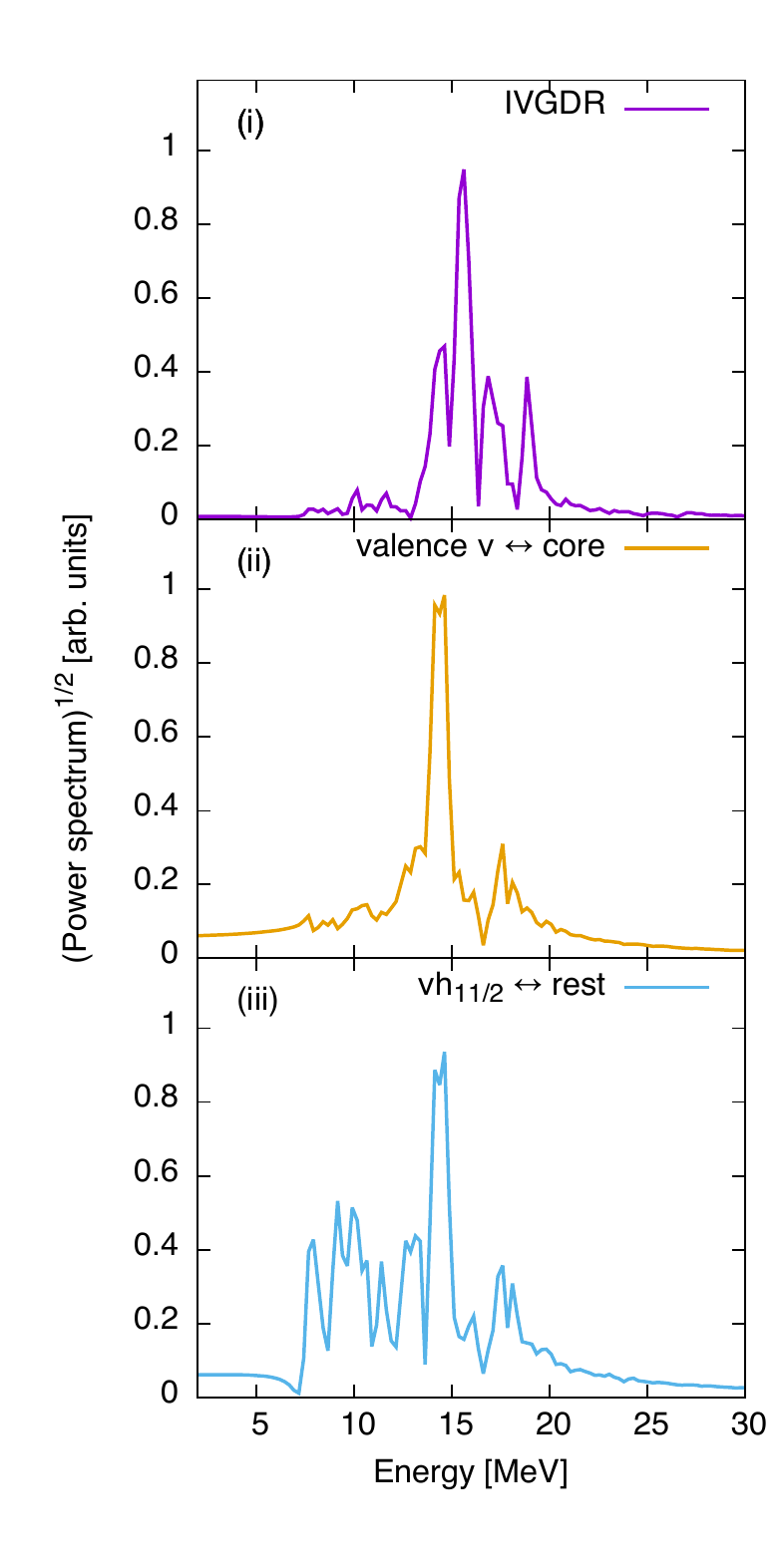}
  \caption{Response of nucleus to three different kinds of external perturbation: (i) a standard isovector dipole kick, (ii) a kick in which all neutrons outside N=50 core are sent in one direction, while all other nucleus are sent in the opposite direction, (iii) a kick in which the $h_{11/2}$ neutrons are sent in one direction, which all other nucleons are sent in the opposite direction}
  \label{sn132power}
\end{figure}

The sums over $i$ and $j$ select which nucleons are boosted in one spatial direaction, and which in the opposite direction.  If $i$ runs over a single nucleon species and $j$ over the other, then the boost is the standard isovector dipole boost in which all protons are set off moving out of phase with all neutrons.  We perform this boost, and also make calculations in which all neutrons above the N=50 ``core'' are set vibrating against the 50 protons and 50 neutrons in this core, and a further calculation in which the least-bound neutron orbital, the $\nu h_{11/2}$, is set in motion against all other nucleons.  Note that in (\ref{eq:boost}) we have chosen explicitly to boost in the $x$ direction since $^{132}$Sn is spherical and any arbitraty boost direction will suffice.

In each of these cases we may expect the response to be at different frequencies.  We follow, as a function of time, the isovector dipole response, i.e. the expectation value of the the operator $\mathbf{D}$ in the case that $i$ runs over neutrons while $j$ runs over protons.  We follow this observable no matter what the boost, and analyse the square root of the Fourier power spectrum.  This gives us information about which modes are excited in the nucleus by each boost, and how those modes couple to the E1 strength.  In the case of the pure IVGDR boost, the spectrum corresponds directly to the strength function \cite{Alm05}.

The resulting excitation spectra are shown in figure \ref{sn132power}.  The top panel, labelled (i), shows the E1 strength, demonstrating a strong GDR peak, between around 13 and 20 MeV, with a smaller low-energy peak between around 7 and 13 MeV.  The spectra are fragmented compared with the experimental results \cite{Adr05} due to the reflecting boundary conditions common to RPA and TDHF calculations \cite{Ste07,Par13,Par14}.

Frame (ii) of figure \ref{sn132power} shows the characteristic vibrational frequencies when one initialises the $2d_{5/2}$, $1g_{9/2}$, $3s_{1/2}$, $2d_{3/2}$ and $1h_{11/2}$ neutrons to be moving with a dipole boost out of phase with all other nucleons.  One sees that the strength mainly lies in the lower part of the giant dipole peak.  The lowest frame, (iii), shows the response frequencies when the $1h_{11/2}$ neutrons are set to vibrate against the rest of the nucleons.  Here, while there is still a strong component in the giant dipole region, the low lying strength is excited with a similar magnitude.  This hints that the low-lying mode has a strong component of skin against core motion, but in which the skin consists of one single orbital.

\begin{figure}[!tbhp]
  \centering
  \includegraphics[width=8cm]{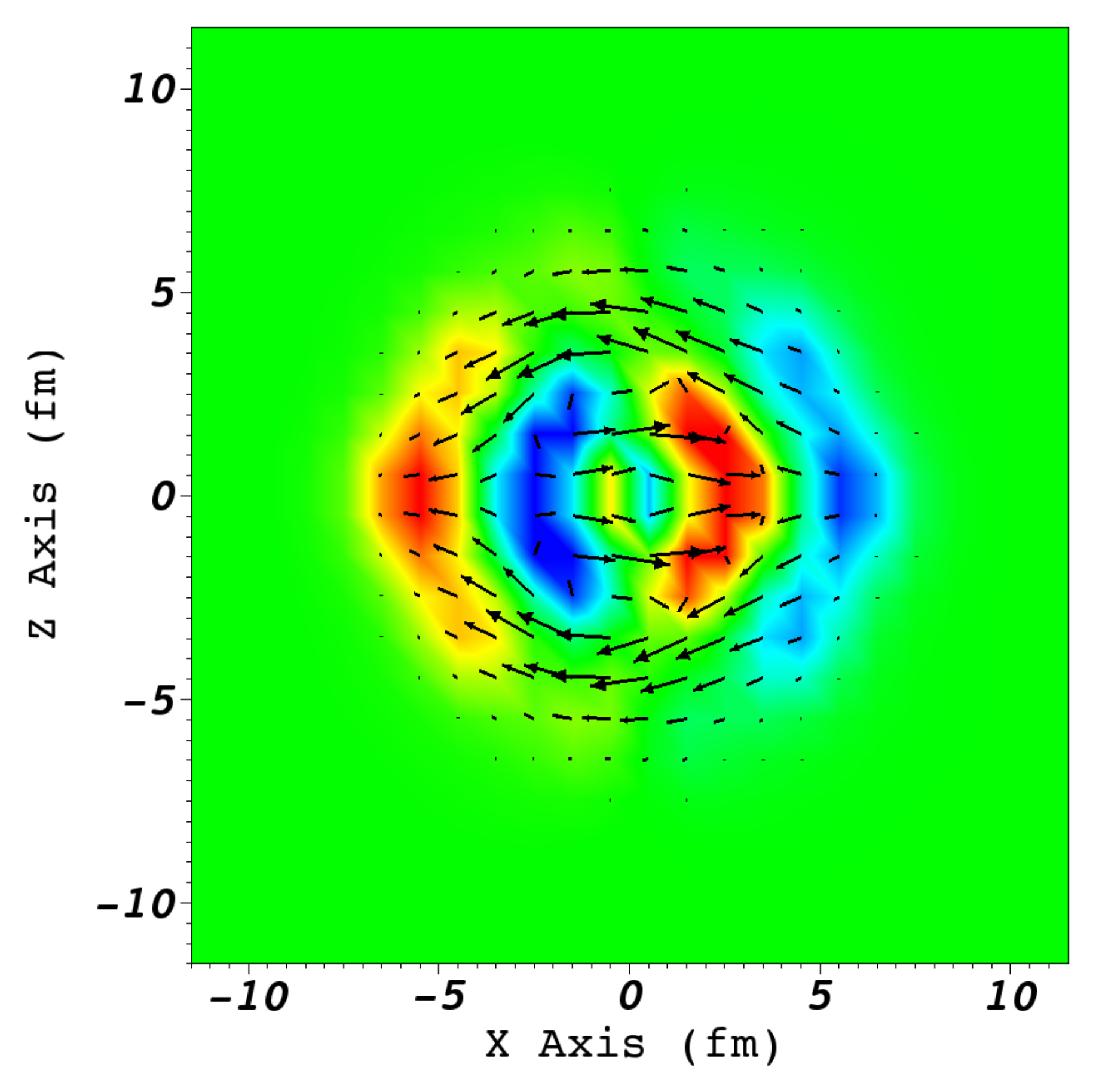}
  \caption{Snapshot, at t=7300 fm/c (24.0 zs).  The colouring indicates the rate of change of neutron density, with red indicating increasing density and blue decreasing.  The arrows show the flow of protons.}
  \label{snapshot}
\end{figure}

A typical snapshot of the dynamic process corresponding to excitation (iii) is shown in figure \ref{snapshot}.  One sees that the protons and neutrons are moving in phase with each other,  since the arrows showing the flow of protons point from the blue regions (in which neutron density is decreasing) to red (in which it is increasing).  The core and the surface are moving out of phase.  This supports the typical picture of the low-lying dipole strength.

\begin{figure*}[!tbhp]
\centering
\includegraphics[width=13.82cm]{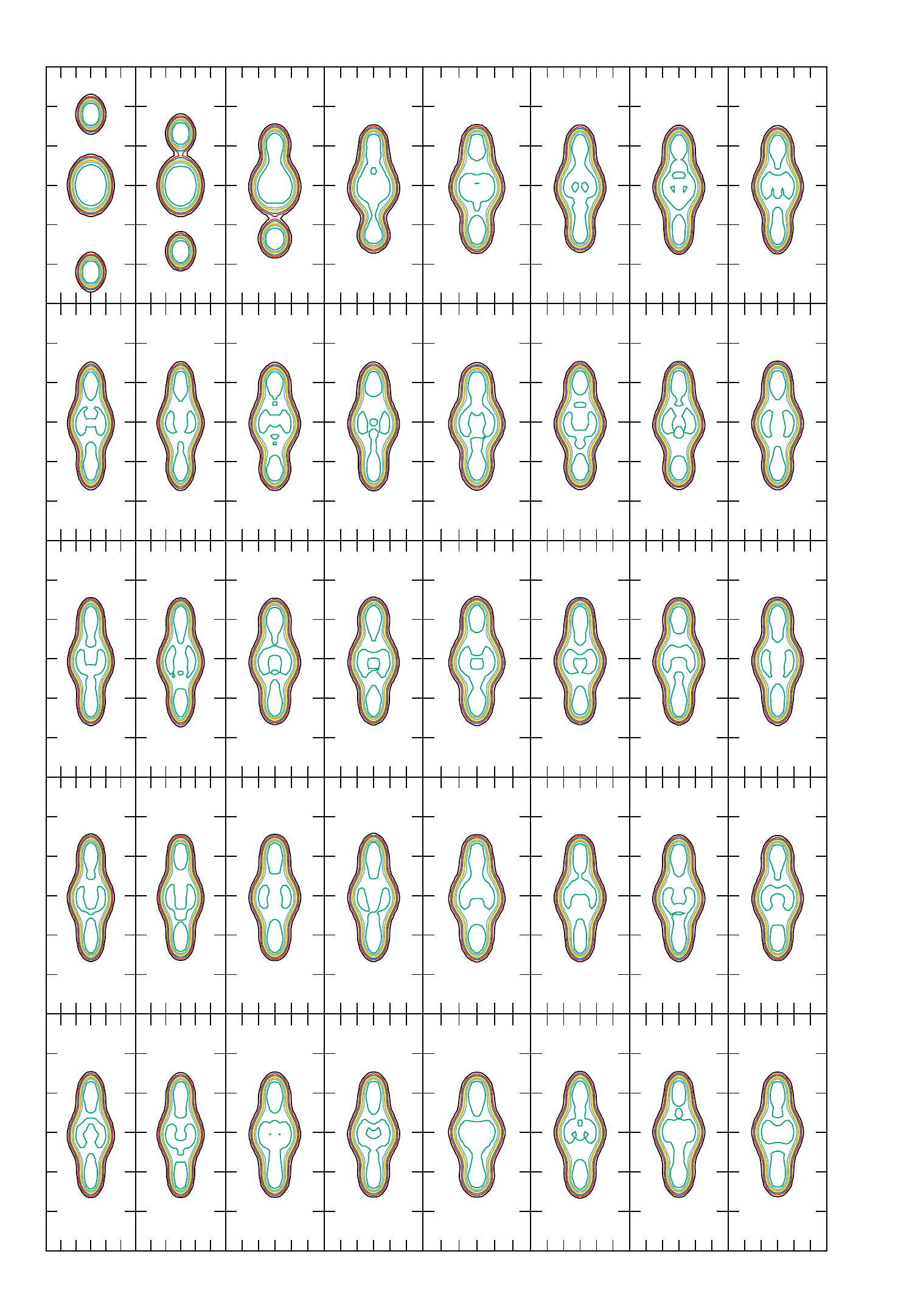}
\vspace*{-0.5cm}       % Give the correct figure height in cm
\caption{Snapshots of the reaction between two $^{48}$Ca nuclei with one $^{208}$Pb nucleus.  Time increases in units of 100 fm/c (0.33 zs) between frames, from left to right and then top to bottom.  Contours show the total density.  Contours are in steps of $0.02$ fm$^{-3}$, with the innermost contour denoting a total density of $0.14$ fm$^{-3}$.}
\label{fusion}       % Give a unique label
\end{figure*}

\section{Ternary Fusion}

Among the many applications of TDHF to nuclear reactions, it has been used to shed light on the dynamics leading to the formation of superheavy elements \cite{UmaSHE}.

For the heaviest elements thus far synthesised, radioactive targets have been used (e.g. a $^{249}$Bk target with a $^{48}$Ca beam for the synthesis of element 117 \cite{Oga10}).  To go much further by the same method, such targets would become increasingly short-lived.  Here, we present a putative method for synthesising superheavy elements involving only stable beams and targets, as well as demonstrating an application of the TDHF technique.  We simulate the two-step reaction
\begin{eqnarray*}
  ^{48}\mathrm{Ca} + ^{208}\mathrm{Pb} &\rightarrow& ^{256}\mathrm{No}^*\\
  ^{48}\mathrm{Ca} + ^{256}\mathrm{No}^* &\rightarrow& ^{304}120
  \end{eqnarray*}

As a sample configuration, we set up an initial condition in which a lead nucleus is placed in the centre of a box of size 24fm $\times$ 24fm $\times$ 60fm, with the two calcium nuclei placed at $(0,0,-22\mathrm{fm})$ and $(0,0,18\mathrm{fm})$.  The two calcium nuclei are given initial kinetic energies of 80 MeV, towards the lead nucleus.  The fact that the calcium nuclei impinge from either side is a calculational convenience, but should be of little consequence since the arrival of the second calcium nucleus should occur on a compound nucleus and would thus be equivalent to a calculation where both calcium nuclei arrive from the same direction.  These fusion calculations are made with the SLy4d Skyrme parameterisation \cite{Kim97}

\begin{figure}[!tbhp]
  \centering
  \includegraphics[width=8cm]{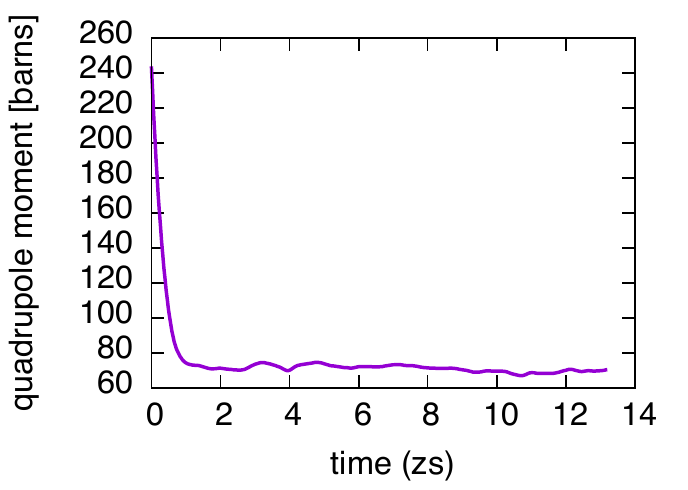}
  \caption{quadrupole moment of system of ternary fusion.}
  \label{quad}
\end{figure}

A series of snapshots of the reaction is shown in figure \ref{fusion}.  The earliest frame is the top left, the latest bottom right and one reads left to right.  The large number of frames in which the nucleus remains in a fused configuration indicates that the nucleus will not undergo e.g. quasi-fission.  Figure \ref{quad} shows the total mass quadrupole moment of the system of the system undergoing reaction.  A the rapid coalescence into a fused single nucleus is seen in the steep initial decrease of the quadrupole moment.  Subsequent motion shows that the nucleus is wobbling, and undergoes many vibrational cycles without fission.

It is also straightforward to find many configurations which do not result in such long-lived fusion.  In the case of ternary fusion there is a large parameter space to map out:  Two impact parameters, and a time delay between the arrival of the two beam nuclei.  Further detailed studies would be needed to produce a more definitive conclusion, and those presented here are indicative only that the method should not be ruled out.  An example of ternary fusion with lighter stable nuclei should in any case be studied as a proof of concept.

\section{Conclusions}
We have presented calculations of giant resonances and fusion reactions, making use of the freedom afforded by the time-dependent Hatree-Fock technique to use quite general starting conditions.  With no further physics input beyond the starting condition, the effective interaction, and the assumption of mean-field dynamics, calculations of resonance modes in $^{132}$Sn have shown that the excitation of the $h_{11/2}$ neutrons against all other nucleons in the system couple to the isovector dipole mode but with an enhanced strength at lower energies.  Futher calculations suggest that a beam of $^{48}$Ca on $^{208}$Pb may be able to produce ternary fusion reactions, though further work is needed to e.g. establish a predicted cross-section.
\section*{Acknowledgements}
Support for this work comes from the UK STFC funding council via grants ST/L005743/1 and ST/J000051/1, along with the award of time on the DiRAC computer system.

\end{document}